\documentclass[8pt]{article}  \usepackage{times}
\usepackage{graphicx}

\usepackage{color}

\usepackage[round,numbers,sort&compress]{natbib} 
\usepackage{amsmath}
\usepackage{amssymb}
\usepackage{stackrel}
\usepackage{chemarrow}

\usepackage{booktabs}
\usepackage{dcolumn}
\usepackage{rotating}
\usepackage{multirow}
\begin{document}

%--- the below commands make the spacing around new sections smaller
%\titlespacing{\section}{0pt}{2pt}{0pt}
%\newenvironment{sciabstract} {\begin{quote} \bf} {\end{quote}}
\twocolumn[{\LARGE \textbf{Nerves and Anesthesia:
A physics perspective on medicine
\\*[0.2cm]}}
{\large T. Heimburg\\*[0.1cm]
{\small Niels Bohr Institute, University of Copenhagen, Blegdamsvej 17, 2100 Copenhagen \O, Denmark}\\
{\small Email: theimbu@nbi.dk}\\*[0.1cm]
{\normalsize ABSTRACT\hspace{0.5cm} We present a recent theory for nerve pulse propagation and anesthesia and argue that both nerve activity and the action of anesthetics can be understood on the basis of simple physical laws. It was found experimentally that biological membranes melt from a solid state to a liquid state just below physiological temperature.  Such melting processes have a profound influence on the physical properties of cell membranes. They make it possible for mechanical pulses (solitons) to travel along nerve axons. In these pulses, a region of solid phase travels in the liquid nerve membrane. These pulses display many properties associated with the action potential in nerves. 
Both general and local anesthetics lower melting temperatures of membranes. Thus, they make it more difficult to excite the nerve membrane. Since hydrostatic pressure increases melting temperatures, it counteracts anesthesia. This theory has the virtue of providing a simple explanation of the famous Meyer-Overton correlation, which states that the effectiveness of an anesthetic is proportional to its solubility in the lipid membranes of cells. We offer evidence that this concept if also applicable to local anesthesia. 
Finally, we show that the presence of transitions has an influence on channel activity that can arise even in the absence of proteins. \\
\\*[0.0cm] }}
]
%------------------------------------------------------------------

\normalsize
%----- INTRODUCTION -----------------------------------------------------------
\begin{minipage}[t]{0.45\textwidth}
\begin{flushright}
\parbox[]{6.0cm}{\textit{ÒThe miracle of the appropriateness of the language of mathematics for the formulation of the laws of physics is a wonderful gift which we neither understand nor deserve. We should be grateful for it and hope that it will remain valid in future research ÉÓ (Eugene Wigner, 1960).  }}\end{flushright}
\end{minipage}

\section*{Introduction}
It is generally believed that the laws of physics, expressed in a mathematical language, are universal \cite{Wigner1960}. They apply to anything ranging from small sub-atomic particles to very large ga\-laxies. While there exist problems where the physical laws are not yet known, there is nevertheless a profound belief that such laws always exist. As pointed out by the physicist Erwin Schr\"o\-dinger in his famous monograph 'What is life?' \cite{Schroedinger1944}, it is to be expected that the laws of physics also govern the function of living matter. The aim of this review is to outline the evidence that some important medical problems, in particular nerve pulse conduction and anesthesia, can be understood on the basis of very simple physical laws.

It is a common misconception that physics has not made significant contributions to biology and medicine. Based on the work of the physicist Max Delbr\"uck, it was Erwin Schr\"odinger who first proposed that DNA is the carrier of genetic information. Delbr\"uckÕs student James Watson together with the physicist Francis Crick found the structure of the DNA helix and proposed a replication mechanism. The atomic physicist George Gamow suggested the genetic code. Max Perutz and John Ken{\-}drew described the first protein structures using x-ray crystallography. It is thus fair to say that physics has contributed materially to the discipline of molecular biology, which now dominates fundamental thinking in medicine. The Òcentral dogmaÓ of molecular biology states that the genes of the DNA encode proteins, and that the proteins display physiological function. It is generally assumed that life is dominated by the function of proteins. Much of drug development in the recent decades focused on the design of particular molecules that block or enhance the activity of proteins, in particular those of channel proteins, ion pumps or certain enzymes. Thus, disease is commonly associated to the function of proteins, and healing is equated to influencing the function of individual proteins or genes.  Indeed, it is tacitly assumed that life can be understood as the sum of the functions of the individual proteins encoded in the DNA. If all proteins are characterized, life seemingly becomes explainable. This is not the view of physics.

%----- end INTRODUCTION -------------------------------------------------------------

%----- begin MATERIALS AND METHODS  -----------------------------------------------------------------
%----- end MATERIALS AND METHODS  -----------------------------------------------------------------

\section*{Size matters}
Physics describes far more than atoms and molecules Ñ it is also successful on larger scales, for example, in describing water waves on a lake, the propagation of sound in air, or the compression of a gas. The characteristic size of waves on water is typically on the order of meters, which is one billion times larger than the size of a water molecule. The mathematical equation that describes the propagation of water waves is extremely simple. In order to understand it, it is not helpful to study the properties of a single water molecule in detail. Waves on alcohol or oil would display similar properties even though their molecular properties are quite different. While the description of a wave appears to be a complicated problem on the atomic scale, its description on a larger macroscopic scale is extremely simple.  Clearly, there are a many simple physical laws that are difficult to understand on a microscopic scale. 
It is generally appreciated in physics that the scale of a theory should match the scale of the phenomena it seeks to explain. Physical properties arising on larger scales are called Òemerging phenomenaÓ. The theories applied to such problems include ÒthermodynamicsÓ and ÒhydrodynamicsÓ. Their primary parameters are temperature, pressure, voltage and chemical potentials, which are associated to entropy, volume, charge and the number of molecules. 
Unfortunately, such theories do not feature prominently in medicine even though they are at the very core of physical understanding. There are numerous examples that demonstrate the value of understanding biology by using physical laws designed to be valid on larger scales. Medical problems that appear difficult on molecular scales may turn out to be more easily understood at larger scales. Here, we will discuss the examples of nerve pulse propagation and anesthesia.

\subsection*{The size of the nervous impulse}
Remarkably, it is not generally appreciated that nerve pulses are quite large. Their size can be calculated as the product of the velocity of the pulse and its duration. For instance, a nerve pulse in a motor neuron has a velocity of about 100 meters per second and duration of 1 millisecond. The resulting length of the nerve pulse is about 10 cm. This is dramatically larger than the scale of the ion channel proteins that are considered fundamental for the explanation of the nervous impulse. The size difference between a potassium channel and the size of a nerve pulse is similar to the difference between a coffee cup and the complete continent of Europe. However, nearly all illustrations of the nerve pulse in neuroscience depict the nerve pulse as something very small (an arbitrary example is shown in Fig.1). This strongly suggests that the neuroscience community has not given much thought to the problem of scales.
%vvvvvvvvvvvvvvvvvvvvvvvvvvvvvvvvvvvvvvvvvvvvvvvvvvvvvvvvvvvvvvvvvvvvvvvvvvvvvvvvvvvvvvvv
\begin{figure}
    \centering
    \includegraphics[width=60mm]{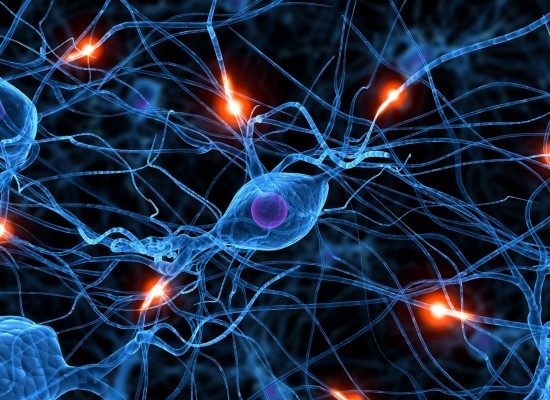}
    \parbox[c]{8.0cm}{ \caption{\textit{The typical drawing of a nervous impulse as a small object is incorrect. Nerve pulses are not microscopic in length, but rather macroscopic \cite{Nerve1}. }
    \label{Figure1}}}
\end{figure}
%^^^^^^^^^^^^^^^^^^^^^^^^^^^^^^^^^^^^^^^^^^^^^^^^^^^^^^^^^^^^^^^^^^^^^^^^^^^^^^^^^^^^^^^^^^^^^^

%========================================================================================
\section*{Phase transitions}
Scales also matter for the melting of ice. Ice melts to water at a temperature of exactly zero degrees.  This process is called a phase transition. Such phase transitions involve the melting of many molecules at the same time, i.e., they are Òcooperative processesÓ. On the surface of a lake at about zero degrees, icebergs and water coexist. Melting is a phenomenon that emerges at a larger scale, and nothing in the structure of a single water molecule indicates that ice should melt at zero degrees. 
%vvvvvvvvvvvvvvvvvvvvvvvvvvvvvvvvvvvvvvvvvvvvvvvvvvvvvvvvvvvvvvvvvvvvvvvvvvvvvvvvvvvvvvvv
\begin{figure}[htb!]
    \centering
    \includegraphics[width=90mm]{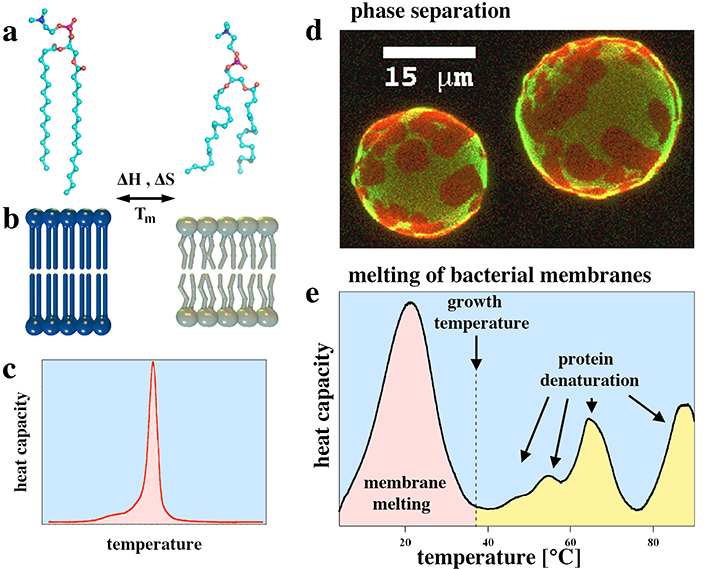}
    \parbox[c]{8.0cm}{ \caption{\textit{Biomembranes consist of lipids and proteins. At low temperatures, lipids are ordered and the membrane exists in solid form (panels a and b, left). At high temperature, the membrane is liquid (panels a and b, right). At the melting temperature, one finds a peak in a calorimetric experiment (panel c). In the melting regime one finds coexistence of solid (red) and liquid (green) membrane domains (panel d, fluorescence microscopy image of a lipid vesicle). The melting point of biological membranes is slightly below their growth or body temperature (panel e; E. coli membranes). Adapted from \cite{Heimburg2009}.}
    \label{Figure2}}}
\end{figure}
%^^^^^^^^^^^^^^^^^^^^^^^^^^^^^^^^^^^^^^^^^^^^^^^^^^^^^^^^^^^^^^^^^^^^^^^^^^^^^^^^^^^^^^^^^^^^^^^
Similar melting processes are found in biological membranes. The lipids of the biomembrane display ordered hydrocarbons chains at low temperature (solid) and disordered chains at high temperature (liquid; Fig. 2, a and b, adapted from \cite{Heimburg2009}). The liquid phase of the membrane has a larger area and a smaller thickness than the solid phase. The transition between these two states of a membrane can be measured in a calorimeter. At the melting temperature, one finds a maximum in the heat capacity profile (Fig. 2c). In the transition, heat is absorbed in order to melt the membrane.  At the transition temperature, well-defined domains of the solid and liquid phases can coexist (Fig. 2d, shown in red and green). The physical properties of membranes change significantly at the phase transition.  Membranes within the transition regime are less rigid, softer (i.e., are more easily compressed), and display larger fluctuations than membranes in the liquid and solid phases.

It is particularly important to realize that the melting point of biomembranes is always a few degrees below physiological temperature. The behavior of E. \textit{coli} membranes shown in Fig. 2e (the lipid melting peak is marked in red) is typical. Similar transitions are found lung surfactant and bacillus \textit{subtilis} membrane \cite{Heimburg2005c} but also in the central nerves of rats and chicken (unpublished). These melting transitions are always found slightly below physiological temperature. If external conditions are changed, the melting temperature slowly adapts to a new growth temperature (E. \textit{coli}), a new hydrostatic pressure (hyperbaric bacteria) or to the continuing presence of solvents such as alcohol \cite{Heimburg2007c}. Melting transitions will play a central role in the physical description of the nerve pulses and anesthesia to be described below. 

\section*{Nerves}
\subsection*{The Hodgkin-Huxley model and the problem of heat production}
The textbook description of nerve pulse conduction models the action potential in terms of currents and voltage changes.   This model was proposed by Alan L. Hodgkin and Andrew F. Huxley in 1952 \cite{Hodgkin1952b}, and they were rewarded with the Nobel Prize in 1963.  The Hodgkin-Huxley model is based on cable theory, which treats the nerve axon as a long cable filled with an electrolyte. Ion currents can flow along the nerve or through the nerve membrane. Hodgkin and Huxley proposed that sodium and potassium currents pass through ion-specific channel proteins called Na-channels and K-channels, respectively. The electrical resistance of these channels is assumed to be a function of voltage, and this property leads to the possibility that a localized pulse of electrical activity can travel along the nerve axon. While the Hodgkin-Huxley model is closer to a physical description than many other models in medical science, it is based on presumptions regarding the function of individual molecules (channel proteins) for which there is no fundamental theory. 
In the framework of this model, proteins are modeled as electrical resistors. It is well known that currents flowing through a resistor generate heat. Imagine a light bulb connected to a battery: It necessarily emits heat as well as light. This emission of heat is independent of the direction of the current.  Currents flowing through a resistor always heat the environment; they never cool it.  According to Hodgkin-Huxley model, the action potential should heat the nerve.  However, this is not what is found in experiments. In fact, it was experimentally shown that the nervous impulse does not lead to net generation of heat \cite{Ritchie1985}. This fact puzzled renowned scientists from Hermann von Helmholtz (who first attempted to measure heat production in nerve in 1848) to Archibald V. Hill (who was rewarded the Nobel Prize in 1922 for heat measurements in Biology) and Alan L. Hodgkin (of the Hodgkin-Huxley model). It is inconsistent with the textbook nerve model. A further indication that current models do not provide a comprehensive description of the action potential can be found in the observation of changes in the thickness and length of nerve axons during the action potential \cite{Iwasa1980a, Iwasa1980b, Tasaki1989}. These experiments suggest that the nerve pulse contains a mechanical component in addition to its more familiar electrical signal.  A satisfactory model of the action potential must account for these mechanical and thermal changes in active nerves.  Existing models simply ignore these important features of the nervous impulse. The absence of net heat generation suggests that the action potential is an adiabatic phenomenon that does not dissipate energy. A familiar example of an adiabatic process is ordinary sound.  We will consider the relation between the propagation of sound and nerve pulses in greater detail below.
%vvvvvvvvvvvvvvvvvvvvvvvvvvvvvvvvvvvvvvvvvvvvvvvvvvvvvvvvvvvvvvvvvvvvvvvvvvvvvvvvvvvvvvvv
\begin{figure*}[t]
    \centering
    \includegraphics[width=120mm]{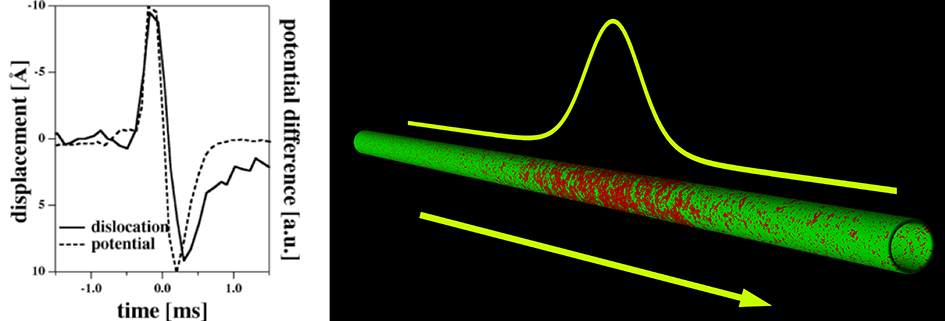}
    \parbox[c]{14.0cm}{ \caption{\textit{The mechanical action potential. Left: During an action potential, the mechanical response of a nerve (solid line) is proportional to the voltage change (dashed line). Adapted from \cite{Iwasa1980a}. Right: The soliton theory predicts a thickness and density change during the action potential. The pulse consists in a solid lipid membrane region (red) traveling in a liquid membrane (green). }
    \label{Figure3}}}
\end{figure*}
%^^^^^^^^^^^^^^^^^^^^^^^^^^^^^^^^^^^^^^^^^^^^^^^^^^^^^^^^^^^^^^^^^^^^^^^^^^^^^^^^^^^^^^^^^^^^^^

\subsection*{Phase transitions and the nervous impulse}
As mentioned, various scientists have noted that nerves change their thickness during the action potential (Fig. 3, left; adapted from \cite{Iwasa1980a}). This change is small but found in all nerves. This suggests the presence of a mechanical signal.  We have shown previously that a biomembrane slightly above its melting point can support electromechanical pulses \cite{Heimburg2005c}. The pulse consists of a compressed (ÒsolidÓ) region traveling in the liquid membrane, and the mechanism of propagation is qualitatively identical to that of sound.  This pulse is called a ÒsolitonÓ. Solitons are localized pulses of compressed membrane that travel unperturbed over long distances. The resulting pulse in membrane density is shown schematically in Fig. 3 (right).

Since the solid membrane is thicker and has a smaller area, the soliton renders the membrane thicker and the nerve contracts \cite{Tasaki1989}. The soliton also exhibits temperature changes similar to those found experimentally \cite{Ritchie1985}. The derivation of the soliton theory involves thermodynamics and hydrodynamics, which is beyond the scope of this review (for technical details see \cite{Heimburg2005c, Lautrup2011}). However, the underlying theory is similar to that describing the propagation of sound, taking into account that the membrane is very compressible in the vicinity of the melting transition. Thus, the soliton can be understood as a sound pulse with electrical properties. This is exactly what is measured in nerves.

\section*{Anesthesia}
In contrast to local anesthesia, it is widely accepted that the origin of general anesthesia is poorly understood.  Rather, most anesthesiologists base their knowledge on practical experience accumulated over decades. However, there are some peculiar findings concerning anesthesia that hint at a generic physical origin of this effect. One of those is the famous Meyer-Overton correlation.
%vvvvvvvvvvvvvvvvvvvvvvvvvvvvvvvvvvvvvvvvvvvvvvvvvvvvvvvvvvvvvvvvvvvvvvvvvvvvvvvvvvvvvvvv
\begin{figure*}[htb!]
    \centering
    \includegraphics[width=12.0cm]{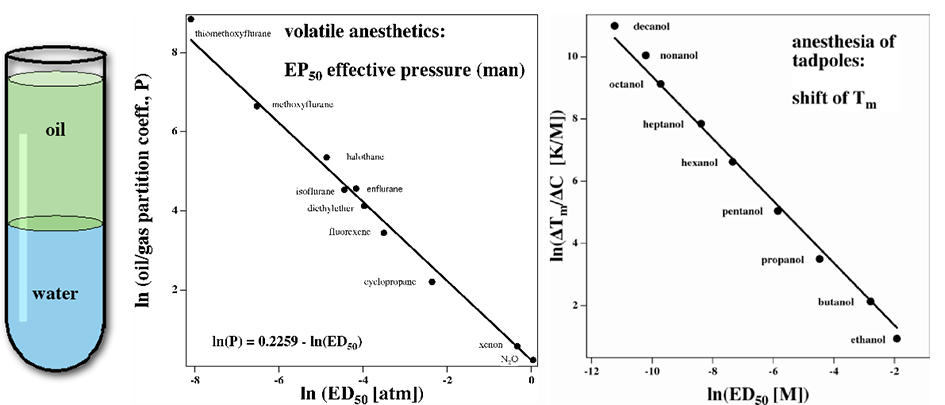}
    \parbox[c]{14.0cm}{ \caption{\textit{Left: The oil-water partition coefficient describes the ratio of anesthetic drug dissolve in oil compared to that dissolved in an equal amount of water. Center: The oil-gas partition coefficient for volatile general anesthetics as a function of critical anesthetic dose, ED50 yields a straight line (adapted from \cite{Overton1991}). Right: The shift in membrane melting temperature, $T_m$, as a function of critical dose, $ED_{50}$, yields a straight line with the same slope. Adapted from \cite{Kharakoz2001}.}
    \label{Figure4}}}
\end{figure*}
%^^^^^^^^^^^^^^^^^^^^^^^^^^^^^^^^^^^^^^^^^^^^^^^^^^^^^^^^^^^^^^^^^^^^^^^^^^^^^^^^^^^^^^^^^^^^^^

\subsection*{The Meyer-Overton correlation}
Imagine adding an anesthetic drug to equal amounts of oil and water (Figure 4, left). One determines how much of the drug dissolves in each of the two phases. The ratio of the two concentrations is called the partition coefficient, $P$. The Meyer-Overton correlation describes the well-known finding that the partition coefficient of general anesthetics between olive oil and water (or air) is inversely proportional to the critical anesthetic dose, $ED_{50}$, at which half of the subjects are anesthetized. This can be expressed as:
\begin{equation}
\label{eq1}
P\cdot [ED]_{50}=constant \;.
\end{equation}
In a double-logarithmic plot of the partition coefficient, $P$, as a function of the critical dose, $ED_{50}$, one expects to find a straight line with a slope of -1, or $\log (P) = \mbox{constant} -$ {\linebreak}$\log ([ED]_{50} )$, see Fig. 4 (center). This correlation is observed over many orders of magnitude for the partition coefficient. The constant in this equation is the same for all general anesthetics ranging from xenon and laughing gas to metoxyflurane.  Its exact value depends on which indicator is used for anesthesia. For tadpole anesthesia one measures the aqueous concentration of drugs when 50\% of the tadpoles sink to the ground (immobility). For humans one takes somewhat different measures. The above correlation was first described by Meyer \cite{Meyer1899} and in a famous book by Overton from 1901 \cite{Overton1901}. It is accompanied by the empirical finding of the additivity of the effects of different anesthetics.  The combination of 50\% of critical dose for each of two anesthetic compounds yields full anesthesia. Overton remarked that the solubility of anesthetics in olive oil basically mimics the solubility within the biomembrane. As also already noted by Overton, this very generic behavior of anesthesia strongly suggests that a very simple physical or physical chemistry law is responsible for this phenomenon. It must fulfill the following two requirements: 1. The effect of an anesthetic is proportional to its dose. 2. The law is absolutely independent of the chemical nature of the anesthetic drug. Unfortunately, Overton did not provide such a law. Thus, the Meyer-Overton finding remained an empirical finding that describes but does not explain anesthesia.

\subsection*{General anesthetics cannot bind to receptors}
There has been a long dispute regarding whether anesthesia is caused by an unspecific effect on membrane lipids or by the specific binding of anesthetic molecules to receptors. In our opinion, the requirements stated above eliminate the possibility that general anesthesia works by a specific binding of the drug to a protein receptor. Xenon is a general anesthetic and simultaneously a chemically inert gas. Thus, it cannot bind specifically to any protein. If the Meyer-Overton correlation is fulfilled, all other general anesthetics must display exactly the same binding properties to a potential receptor as xenon. Since xenon cannot associate with a receptor, it is likely that other general anesthetics cannot bind to receptors either.
%vvvvvvvvvvvvvvvvvvvvvvvvvvvvvvvvvvvvvvvvvvvvvvvvvvvvvvvvvvvvvvvvvvvvvvvvvvvvvvvvvvvvvvvv
\begin{figure*}[htb!]
    \centering
    \includegraphics[width=12.0cm]{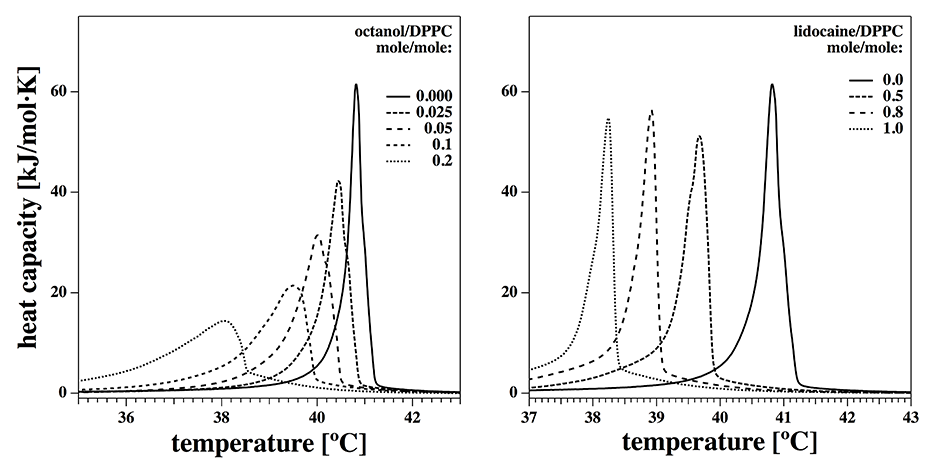}
    \parbox[c]{14.0cm}{ \caption{\textit{Both, general and local anesthetics shift the melting temperature, $T_m$, of the synthetic lipid DPPC towards lower temperatures. Left: Shift in $T_m$ for the general anesthetic octanol at five concentrations. Right: Shift in $T_m$ for the local anesthetic lidocaine at four concentrations. From \cite{Graesboll2014}.}
    \label{Figure5}}}
\end{figure*}
%^^^^^^^^^^^^^^^^^^^^^^^^^^^^^^^^^^^^^^^^^^^^^^^^^^^^^^^^^^^^^^^^^^^^^^^^^^^^^^^^^^^^^^^^^^^^^^

\subsection*{Phase transitions and anesthesia}
Various authors have shown that the presence of anesthetics lowers the melting temperature of membranes. (e.g., \cite{Trudell1975, Heimburg2007c, Graesboll2014} and references therein). This is true for both general and local anesthetics. Fig. 5 shows this for the examples of the general anesthetic octanol and the local anesthetic lidocaine in an artificial membrane. It was found that the lowering of the melting temperature, $\Delta T_m$ (i.e., shifting the peak shown in Fig. 2 c to the left) is described accurately by the simple equation
\begin{equation}
\label{eq2}
\Delta T_m=-\frac{R T_m^2}{\Delta H}x_A \;,
\end{equation}
where $x_A$ is the molar fraction of anesthetic in the membrane. Here, $R$ is the gas constant, and $\Delta H$ is the heat of melting of the membrane (both are constants). The underlying assumption is that anesthetics are ideally miscible in liquid membranes and absolutely immiscible in solid membranes. This law originates from the Dutch physical chemist Jacobus Henricus vanÕt Hoff. It has been known since 1886 as the freezing-point depression law. Originally, it described the lowering of the melting temperature of ice by salt (as we observe it every winter on icy roads). Thus, anesthetics act as anti-freeze agents in membranes. The molar fraction of anesthetics in the membrane is proportional to the partition coefficient. For this reason, it is to be expected that the effect of anesthetics on the melting temperature is inversely proportional to the critical dose, $ED_{50}$. This is shown for a series of alcohols in Figure 4 (right, adapted from \cite{Kharakoz2001}). Moreover, it is immediately apparent from eq. (2) that the depression of the freezing point is proportional to the molar concentration of anesthetics in the membrane and that it is absolutely unspecific with respect to the chemical nature of the anesthetic drug. These two properties naturally lead to the required additivity of the effect of two drugs. Thus, eq. (2) has exactly the necessary properties that follow from the Meyer-Overton correlation.

Interestingly, the effect of local anesthetics on the melting temperature obeys the same theoretical description as the effect of general anesthetics  \cite{Graesboll2014}. Thus, the thermodynamic effect of general and local anesthetics is the same. Unfortunately, a correlation such as the Meyer-Overton correlation is not known for local anesthetics. A critical dose is difficult to determine because intravenous injection can cause cardiac arrest. However, there are a few studies (e.g., after incidental intravenous injection of local anesthetics) that show that the effects of general anesthetics and local anesthetics are additive. General anesthesia can be induced at lower concentrations when local anesthetics are present in the blood at low doses (discussed in  \cite{Graesboll2014}). This indicates that local anesthetics also have the characteristics of a general anesthetic. Further, the shift in melting temperature is inversely proportional to the partition coefficient of local anesthetics. This hints indirectly that the Meyer-Overton correlation might also be valid for local anesthetics. It is therefore plausible to postulate that the physical mechanism behind local anesthesia is similar to the one explaining general anesthesia.

\subsection*{Pressure reversal of anesthesia}
It could be that the relation between melting temperature and critical dose described above is merely a correlation and does not imply a causal relation. It hints at that melting temperatures of biological membranes are related to anesthesia but it does not prove it, and it does not explain why melting is important. Independent evidence for the relevance of melting points is provided by the interesting phenomenon of pressure reversal of anesthesia. It was shown by Johnson and Flagler \cite{Johnson1950} that anesthetized tadpoles wake up at pressures above approximately 50 bar.  Since the solid phase of lipids has a higher density than the liquid phase, pressure also increases the melting temperature of membranes \cite{Trudell1975, Ebel2001}. Thus, if anesthetics lower the melting temperature and pressure increases it, it is expected that the two effects cancel at a given pressure that can be calculated. For the values of artificial membranes one calculates that pressure reversal of general anesthesia in tadpoles should occur for pressure in excess of approximately 25 bar. This is close to the measured values. 
Pressure reversal thus supports the notion that the melting temperature of biomembranes is related to anesthesia.

\subsection*{Reversal by pH and inflammation}
A similar effect can be found for pH changes. Lowering the pH increases the melting temperature of membranes \cite{Heimburg2007c}. Thus, a decrease in pH should be antagonistic to anesthesia. It is known that inflamed tissue displays a lower pH (about 0.5 pH units \cite{Punnia-Moorthy1987}. In fact, it has been found that tissue with inflammations cannot be anesthetized at the same dose. The pH-effect is sufficient to explain the reversal of anesthesia by inflammation (see \cite{Heimburg2007c}). While some local anesthetics display pH dependence themselves (e.g., lidocaine), this should hold also for anesthetics that are not pH sensitive.

\subsection*{Cutoff-effect of long-chain alcohols}
All alcohols up to a chain-length of 12 (dodecanol) act as general anesthetics. As shown in Fig. 4 (right), they also cause melting-point depression following eq. (2). However, longer-chain alcohols fail to be general anesthetics \cite{Pringle1981}. This finding has often been used as an argument against the Meyer-Overton correlation because long-chain alcohols dissolve in membranes but do not cause anesthesia. Therefore, not everything that dissolves in membranes causes anesthesia. However, it has been shown that the long-chain alcohols with chain-lengths of 14 and higher also do not cause melting-point depression \cite{Kaminoh1992}. Thus, the cutoff-effect is in agreement with the melting-temperature depression mechanism.
%vvvvvvvvvvvvvvvvvvvvvvvvvvvvvvvvvvvvvvvvvvvvvvvvvvvvvvvvvvvvvvvvvvvvvvvvvvvvvvvvvvvvvvvv
\begin{figure*}[htb!]
    \centering
    \includegraphics[width=12.0cm]{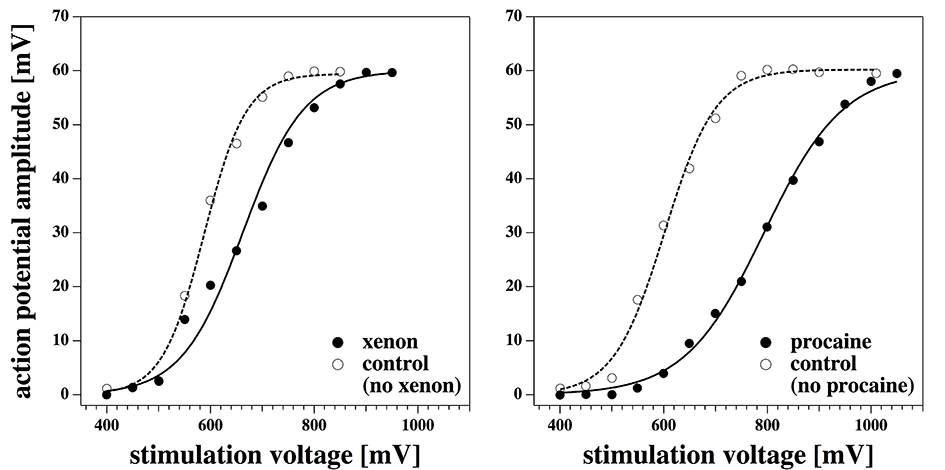}
    \parbox[c]{14.0cm}{ \caption{\textit{Stimulus-response curves for compound action potential in the sciatic nerve of frogs. Left: In the presence and absence of the general anesthetic xenon. Right: in the presence and absence of the local anesthetic procaine. The influence of general and local anesthetics on stimulus-response curves is similar. Adapted from \cite{Kassahun2010}.}}}
    \label{Figure6}
\end{figure*}
%^^^^^^^^^^^^^^^^^^^^^^^^^^^^^^^^^^^^^^^^^^^^^^^^^^^^^^^^^^^^^^^^^^^^^^^^^^^^^^^^^^^^^^^^^^^^^^

\subsection*{Summary of anesthetic effect}
As shown above, the many known anesthetic phenomena are in agreement with melting-point depression caused by anesthetics. The Meyer-Overton correlation is a valid but purely empirical correlation that does not explain anesthesia. However, melting-point depression is a true physical explanation of the anesthetic effect because it influences nerve activity as will be shown below. For most general anesthetics, both correlations are in agreement. However, the melting-point depression also explains pressure reversal, the effect of inflammation and the cutoff effect. Thus, it is a powerful tool that should be of value in the search for new anesthetics. Everything that lowers melting temperatures should display anesthetic effects. Everything that raises melting temperatures is antagonistic. Further, it seems likely that local anesthetics have many characteristics in common with general anesthetics.

\subsection*{Stimulus-Response curves}
Freezing-point depression plays a distinct role in the soliton theory for nerve pulse propagation. We have shown that the free energy (the stimulus) necessary to initiate a soliton pulse in nerves depends on the distance of physiological temperature from the melting temperature of the cell membranes. The larger this distance, the greater the stimulation required to trigger nerve activity. Anesthetics lower the melting temperature of biomembranes and thereby increase the stimulation threshold for nerve pulses.

Kassahun et al. \cite{Kassahun2010} demonstrated that both the general anesthetic xenon and the local anesthetic procaine increase the stimulation threshold of the sciatic nerve of frogs in agreement with the above predictions (Fig. 6). Thus, general anesthesia slightly increases the stimulation threshold but it does not render the nerves inactivate. This may be sufficient to inhibit cooperative higher brain functions such as conscience while basic nerve functions are mostly maintained. Further evidence was provided by \cite{Moldovan2014} who show that the stimulation threshold can change by more than a factor of 10 upon application of the local anesthetic lidocaine to the median nerve in humans. Shifts in threshold of this magnitude are impossible to explain by the Hodgkin-Huxley model or related models.

\subsection*{General and local anesthetics}
The similarity of the effect of general and local anesthetics is surprising because it is widely assumed that the two classes of drugs act by different mechanisms. While it is agreed upon that general anesthesia is poorly understood, local anesthesia is thought to be associated with the blocking of ion channel proteins Ñ in particular to the inhibition of sodium channels \cite{Butterworth1990}. Those channels play an important role in the Hodgkin-Huxley model for nerve pulse propagation (see above). The apparent blocking of channels by local anesthetics has been demonstrated in numerous patch-clamp experiments. While this would seem to settle the issue, the interpretation of the data is neither obvious not clear.

\section*{Ion channels}
Ion channels are typically observed in patch-clamp experiments. A tiny glass pipette is attached to a cell surface, and a slight suction pressure is applied. Inside of the pipette a little electrode is located. The counter-electrode is placed inside the cell. If a voltage is applied, one can measure small electrical currents through the cell membrane. These currents are not constant, but rather display distinct on-off characteristics that have been associated with the opening and closing of ion channel proteins. 

\subsection*{Phase transition and lipid ion channels}
We and other groups have shown that membranes can display ion-channel-like current traces even in the complete absence of proteins provided only that the membrane is close to a melting transition (Fig. 6; \cite{Blicher2009, Laub2012, Blicher2013}). These events are called Ôlipid channelsÕ. Lipid channels consist of small pores (or defects) in the lipid membrane that open and close. 
In patch-clamp experiments, protein channel traces and recordings from synthetic membranes completely devoid of proteins are literally indistinguishable \cite{Laub2012}. Like protein channels, lipid channels are voltage-gated (Fig. 6), mechanosensitive, and can be affected by calcium \cite{Heimburg2010}. 
Lipid channels are more likely in the melting regime of membranes because density fluctuations in transition regions are far more frequent. Therefore, thermal motion is sufficient to open pores in the membrane. Consequently, lipid channels can be blocked by anesthetics due to their effect on melting transitions \cite{Blicher2009}. For the general anesthetic octanol, the blocking of the Na-channel and the acetylcholine receptor displays very similar characteristics than the blocking of lipid channels. While there exists clear evidence that lipid membranes in the absence of proteins display channel events in patch-clamp experiments, it is difficult to prove that ion current actually passes through a protein channel because one cannot record protein properties at nanometer scales or in the absence of a lipid membrane. 
%vvvvvvvvvvvvvvvvvvvvvvvvvvvvvvvvvvvvvvvvvvvvvvvvvvvvvvvvvvvvvvvvvvvvvvvvvvvvvvvvvvvvvvvv
\begin{figure}[htb!]
    \centering
    \includegraphics[width=9.0cm]{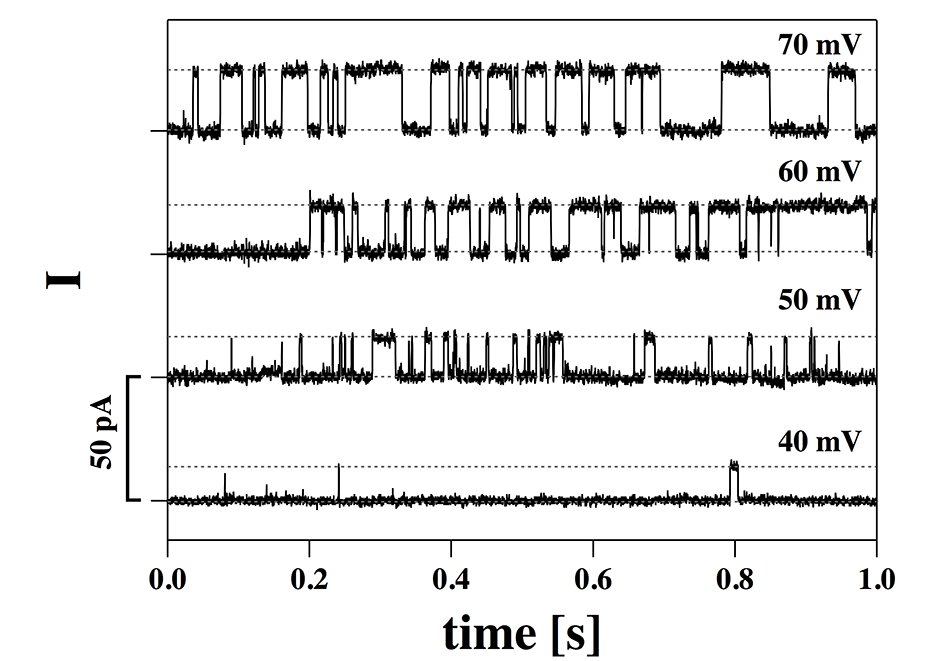}
    \parbox[c]{8.0cm}{ \caption{\textit{Close to the lipid melting temperature of DPPC in the complete absence of proteins one finds ion-channel-like ion conduction events in patch-clamp experiments, which are voltage-gated. The signature is indistinguishable from those attributed to protein ion channels. Shown are single-channel conduction traces at four voltages. These lipid channels are also blocked by anesthetics. From \cite{Blicher2013}.}
    \label{Figure7}}}
\end{figure}
%^^^^^^^^^^^^^^^^^^^^^^^^^^^^^^^^^^^^^^^^^^^^^^^^^^^^^^^^^^^^^^^^^^^^^^^^^^^^^^^^^^^^^^^^^^^^^^

What one can learn from these findings is \cite{Blicher2009}:
\begin{enumerate}
  \item The recording of channel events in biological membranes does not prove the activity of a channel protein. These events could originate from the lipid membrane.
  \item The reduction of the number of channel events by anesthetics does not prove that a channel protein is blocked. The blocking may be due to an absolutely unspecific effect of the drugs on the physical properties of the overall membrane (such as melting-point depression).
For this reason, we have proposed that some proteins may actually not be channels themselves but rather catalysts for lipid channels \cite{Mosgaard2013}. The influence of drugs on ion channels in many cases may well be a generic physical effect that is not directly related to proteins.
\end{enumerate}	
%========================================================================================

\section*{Summary}
The melting of membranes is a process that cannot be understood at the level of single molecules. Features of such transitions become apparent only on mesoscopic or macroscopic scales. We have shown here that it is possible to understand nerve pulses and anesthesia as well as channel activity on the basis of melting processes. In particular, the presence of transitions allows for electromechanical density pulses to travel along nerve axons. Anesthetics lower the transition temperature and thereby increase the threshold for nerve stimulation. Pressure increases transition temperatures and thereby reverses anesthesia. It is interesting to note that many changes in macroscopic parameters can change melting points and thereby potentially influence anesthesia. For instance, it is to be expected the lowering pH and increasing calcium concentration will antagonize anesthesia, while increasing pH causes anesthesia. A\-nesthesia should be sensitive to changes in body temperature, pressure or ion concentrations in general.
Transitions exist also in other biomaterials, e.g., in the protein networks on cell surfaces, in DNA and in large proteins. A lesson from our findings is that one should avoid searching for the origin of medical phenomena exclusively on molecular scales. Surely, molecular medicine has been very important for treating diseases, but there exists a plethora of conditions of a body that cannot be understood on the level of individual molecules. Thermodynamics is a theory of molecular ensembles. When applying it to living organisms it suggests the consideration of the total cell membrane or even a complete cell. On that scale, laws may actually be simpler than on the molecular level. Due to slight changes in a variable (e.g., pH), many biological functions of cells could change without that a discrete molecular origin can be identified. Nevertheless, there may still exist very simple explanations as demonstrated here for nerve pulses, anesthesia and the blocking of channels.

\vspace{0.5cm}
%----- begin ACKNOWLEDGEMENTS ---------------------------------------
\textbf{Acknowledgements: } I thank Prof. Andrew D. Jackson from the Niels Bohr International Academy in Copenhagen for a critical reading of the manuscript. This work was supported by the Villum Foundation (VKR 022130).\\
%----- end ACKNOWLEDGEMENTS -----------------------------------------

\small{
%\bibliography{/Users/thomasheimburg/Documents/0_Thomas/TH_MSBerichteAntr/Manuskripte/Bibtex_central/Bibdesk_central_literature}

}

\end{document}